\documentclass{article}

\usepackage{geometry}
\usepackage{doc}

\usepackage{url}

\usepackage{graphicx}
\usepackage{epstopdf}
\usepackage{hyperref}
\usepackage{mathrsfs}
\usepackage[ansinew]{inputenc}
\usepackage{float}
\usepackage{afterpage}
\usepackage{authblk}
\usepackage{amsmath}

\title{Machine learning assisted droplet trajectories extraction in dense 
emulsions and their analysis}

\author[1]{Mihir Durve}
\author[2]{Adriano Tiribocchi}
\author[3]{Andrea Montessori}
\author[2]{Marco Lauricella}
\author[1,2,4]{Sauro Succi}

\affil[1]{Center for Life Nano- \& Neuro-Science, Fondazione 
Istituto Italiano di Tecnologia (IIT), viale Regina Elena 295, 00161 Rome, 
Italy}
\affil[2]{Istituto per le Applicazioni del Calcolo del Consiglio 
Nazionale delle Ricerche, via dei Taurini 19, 00185, Rome, Italy}
\affil[3]{Dipartimento di Ingegneria, Università degli Studi Roma tre, via 
Vito Volterra 62, Rome, 00146, Italy
}
\affil[4]{ Department of Physics, Harvard University, 17 Oxford St, Cambridge, 
MA 02138, United States
}

\date{}

\begin{document}

\maketitle

\abstract{This work analyzes trajectories obtained by YOLO and DeepSORT 
algorithms 
of dense emulsion systems simulated by Lattice Boltzmann methods. The results 
indicate that the individual droplet's moving direction is 
influenced more by the droplets immediately behind it than the droplets in 
front 
of it. The analysis also provide hints on constraints on writing down a 
dynamical model of droplets for the dense emulsion in narrow channels. 

}





%
%

\section{Introduction}

The last decades have witnessed an impressive rise of machine learning methods, 
which have profoundly impacted many areas of science, ranging from high energy 
physics and quantum computation to material design, and have played a vital 
role 
in recent milestone discoveries~\cite{Alfa_Fold, CARLEO}. Many machine 
learning-based applications are targeted to perform specific tasks such as 
image 
recognition, natural language processing,  sentiments analysis, and  
handwriting 
recognition, to name a  few~\cite{darmatasia,tandel,ahlawat,han2014speech}, 
with 
the aim of improving accuracy, decreasing processing speed, and reducing human 
efforts to perform tasks that are perceived as mundane or labor intensive. In 
microfluidics, in particular, they have been used to study shapes and predict 
transport properties of flowing droplets~\cite{hadikhani,mahdi,khor}. 

Deep neural networks have become one of the fundamental mathematical models for 
implementing machine learning algorithms~\cite{goodfellow}.
Computer vision is one such application domain that uses deep neural networks to 
perform essentially two tasks, object recognition and object tracking. For 
example, a computer vision application connected to a  camera observing a 
traffic junction could easily count the number of cars passing through it and 
monitor their speed, thus making it a widespread tool for smart 
traffic management~\cite{osman2017}.

Recently, we employed two state-of-the-art computer vision algorithms, namely 
You Only Look Once (YOLO) and DeepSORT, to recognize and 
track droplets in high internal phase dense emulsions and infer 
individual droplets trajectories. These emulsions consist of highly ordered 
liquid droplets arranged in crystal-like 
structures~\cite{montessori2021_1,montessori2021_2,bogdan2022}, and have shown 
promising  applications in electrochemical sensing and tissue engineering 
\cite{costantini2014}. 
Such materials represent a fundamental challenge to  non-equilibrium 
thermodynamics as they feature highly non-Newtonian mechanical and rheological 
properties. Thus, studying the rich dynamics of these classical 
many-body systems is crucial to optimize their disegn as well as the 
microfluidic devices employed for their synthesis and applications.

In this work, we revisit the tools employed to infer the trajectories of the 
droplets in dense flowing emulsions simulated using lattice Boltzmann 
methods, presenting an analysis of such trajectories (obtained via computer 
vision tools, see Fig.10 and video4.avi in Ref.\cite{durve_epjp}) in terms of 
quantities generally adopted to characterize the behavior of active matter 
systems.

The paper is structured as follows.
In the next Section we describe the system under investigation; later in 
Section~\ref{sec:alorithms}, we take a brief overview of the computer vision 
algorithms deployed to get individual droplet trajectories and in 
Section~\ref{result} 
we present the analysis of the inferred trajectories. 

\section{Physical system}
\label{sec:LB}

\begin{figure}[]
\centering
\includegraphics[scale=0.9]{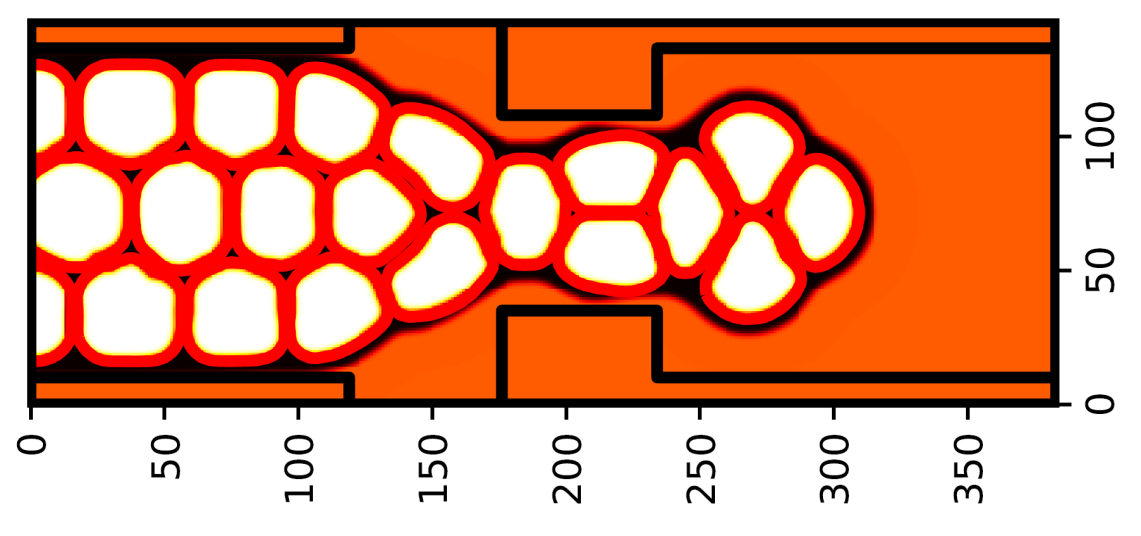}
\caption{\label{fig:translocation} Snapshot of a dense emulsion 
simulated by Lattice Boltzmann methods.}
\end{figure}

The system under study is a soft granular material made of approximately 
monodisperse fluid droplets (white region in Fig.~\ref{fig:translocation}) 
immersed within an inter-droplet continuous phase (dark orange lines) and 
surrounded 
by an external bulk phase (region outside the emulsion). Its structure closely 
resembles that of a double emulsion with multi-core 
morphology~\cite{utada2005,costantini2014,montessori2021_1}, where a high volume 
fraction of dispersed droplets, generally above the close packing limit for 
spheres, arrange in a tightly packed configuration. This material is produced 
within a microfluidic channel which, in our system, consists of an inlet  
reservoir followed by a thinner channel connected to a further downstream 
reservoir. Further details about the simulations, performed via lattice 
Boltzmann methods \cite{montessori2015lattice,coreixas2019comprehensive, 
succi2018lattice}, can be found in Ref.~\cite{durve_ptrsa}.

In the next section, we briefly describe the algorithms employed to analyze 
videos of the physical system.

\section{Algorithms}
\label{sec:alorithms}

Two algorithms were combined to achieve droplet tracking. The first one, 
called You only look once (YOLO), is tasked with identifying the 
droplets in an input image while the second one, called DeepSORT, for 
tracking the droplets in sequential frames.

\subsection{You Only Look Once (YOLO)}

YOLO is a single-stage state-of-the-art object detection 
algorithm. The current version of YOLO (YOLOv5) is the 
fastest and most accurate object detector on two commonly used, general-purpose 
object detection datasets called Pascal VOC (visual object 
classes)~\cite{pascal} and Microsoft COCO object detection datasets~\cite{coco}. 
The image analysis speed, i.e. inference speed of the YOLOv5 networks, is at or 
above  60 frames per second (FPS)~\cite{zhou2021, luiz2021} for general object 
detection.

The YOLO algorithm is the fastest due to its smart operating 
procedure~\cite{redmon, redmon1}. The input is divided in a $S \times S$ grid,
with each cell responsible for detecting an object within
the cell. Each grid cell then predicts $B$ bounding boxes with their 
confidence score for each detected object and $C$ conditional class 
probabilities for the given object belonging to a specific class. This 
information is then combined to produce the final output 
as a single bounding box around the detected object and the class of that 
object. This final output is then passed to the object tracking DeepSORT 
algorithm.

The training dataset is used to train the YOLO networks for recognizing the 
droplets. The training dataset contains several images of droplets and 
associated label files, which include the location and dimensions of the 
droplets in each image. The training data is used to get predictions and update 
the network parameters based on true and predicted output in an iterative 
process. The technical details of the training procedure to train a YOLO 
network 
for droplet recognition is described here~\cite{durve_epjp, durve_ptrsa}.

\subsection{DeepSORT}

The DeepSORT algorithm 
constructs trajectories of all the detected objects by analyzing sequential 
frames \cite{wojke}, employing a classical Simple Online 
Real-Time tracking module~\cite{Bewley2016_sort} in the first stage. This 
module uses the Hungarian 
algorithm~\cite{hungarian} to distinguish detected objects in two 
consecutive frames and assigns individual objects their unique identity. The 
module also uses a Kalman filtering~\cite{kalman} for predicting the 
future position of the objects based on their current positions. 
At the second level, the deep network 
learns object descriptor features to minimize the identity switches as the 
object moves in subsequent frames. The YOLO and the DeepSORT algorithms 
together accomplish droplet recognition and tracking, thus allowing to obtain 
the trajectories of individual droplets (see Fig.10 in Ref.~\cite{durve_epjp}). 
These ones will be analyzed in the next section. 

\section{Analysis of the droplet trajectories}
\label{result}

Preliminary observations (see video4.avi in Ref.~\cite{durve_epjp}) indicate 
that, in a densely packed configuration, the droplets are moving along the same 
direction, in a way akin to birds in highly polar 
flocks~\cite{Cavagna2015,ballerini}. To dig more deeply into such analogy, we 
compute a polar order parameter $\psi$ which is typically used to measure 
direction consensus in moving particles~\cite{vicsek1995}. It is  computed as
\begin{equation}
 \psi = \frac{1}{N} \bigg\lvert \sum_{i=1}^{N} \frac{{\bf v}_i}{\vert {\bf v}_i 
\vert} \bigg\rvert,
\end{equation}

were N is the total number of droplets, ${\bf v}_i$ is the velocity vector of 
the $i^\text{th}$ droplet and $\vert . \vert$ is the modulus of a vector. If 
$\psi=1$ all the droplets move in the same 
direction while $\psi \approx 0$ means that droplets move randomly.  

\begin{figure} []
\centering
\includegraphics[scale=0.5]{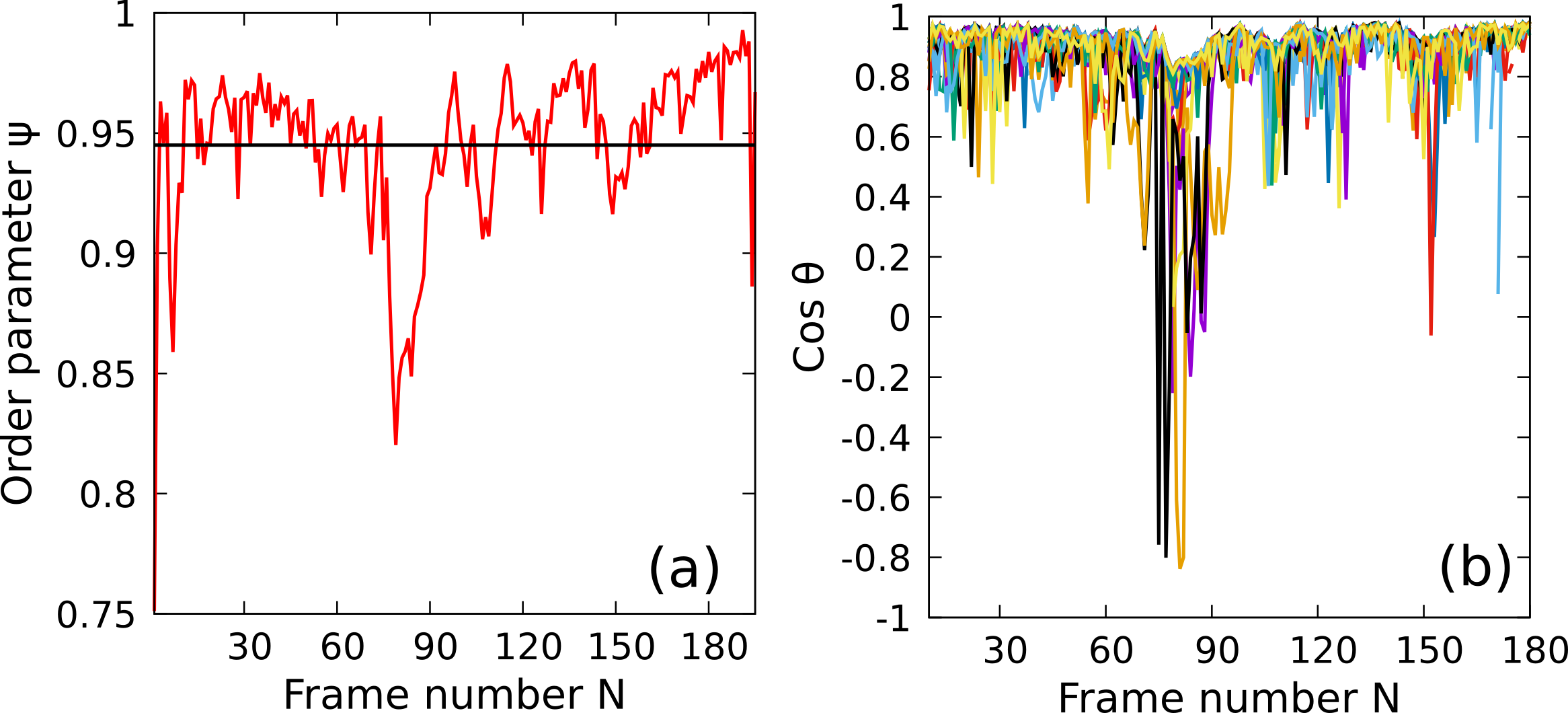}
\caption{(a) Polar order parameter $\psi$ computed at each 
frame. It accounts for moving direction consensus. The black line indicates 
average of $\psi$ over all the frames. (b) Cosine of angle 
between individual droplet's moving direction and average 
direction of all the droplets. Cosine of the angle computed for individual 
droplet is shown with a unique color.\label{order_para}} 
\end{figure}

The polar order parameter is shown in Fig.~\ref{order_para}(a). The high 
average 
value of $\psi$ ($\psi>0.94$) indicates that the droplets are moving more or 
less in the same direction except at a time interval when they are about to 
enter the 
narrow channel. In Fig.~\ref{order_para}(b), we compare an individual droplet's 
heading direction with the average moving direction of the emulsion. This 
comparison is made by measuring the cosine of an angle $\theta$ between the 
velocity vector of the emulsion (taken as the resultant velocity vector of all 
the droplets) and the individual droplet's velocity vector. Except before 
entering the narrow channel, the droplets move in the direction of their bulk 
motion as indicated by the cosine value close to unity. In a system containing 
self-propelled particles (flock of birds, for example), such a high value of 
polar order parameter and cosine of $\theta$ corresponds to a highly ordered 
state 
in which the agents move in a common direction~\cite{Cavagna2015}. These 
observations show apparent 
similarities between a flock of birds and the emulsion in terms of dynamical 
properties. These similarities raise an interesting question, i.e.  whether the 
well-known  models~\cite{vicsek1995,couzin2002,barberis} to mimic the behavior 
of active  matter systems by taking into account 
only local interactions can account for the dynamical properties of the 
emulsion observed here.

\begin{figure} []
\centering
\includegraphics[scale=0.7]{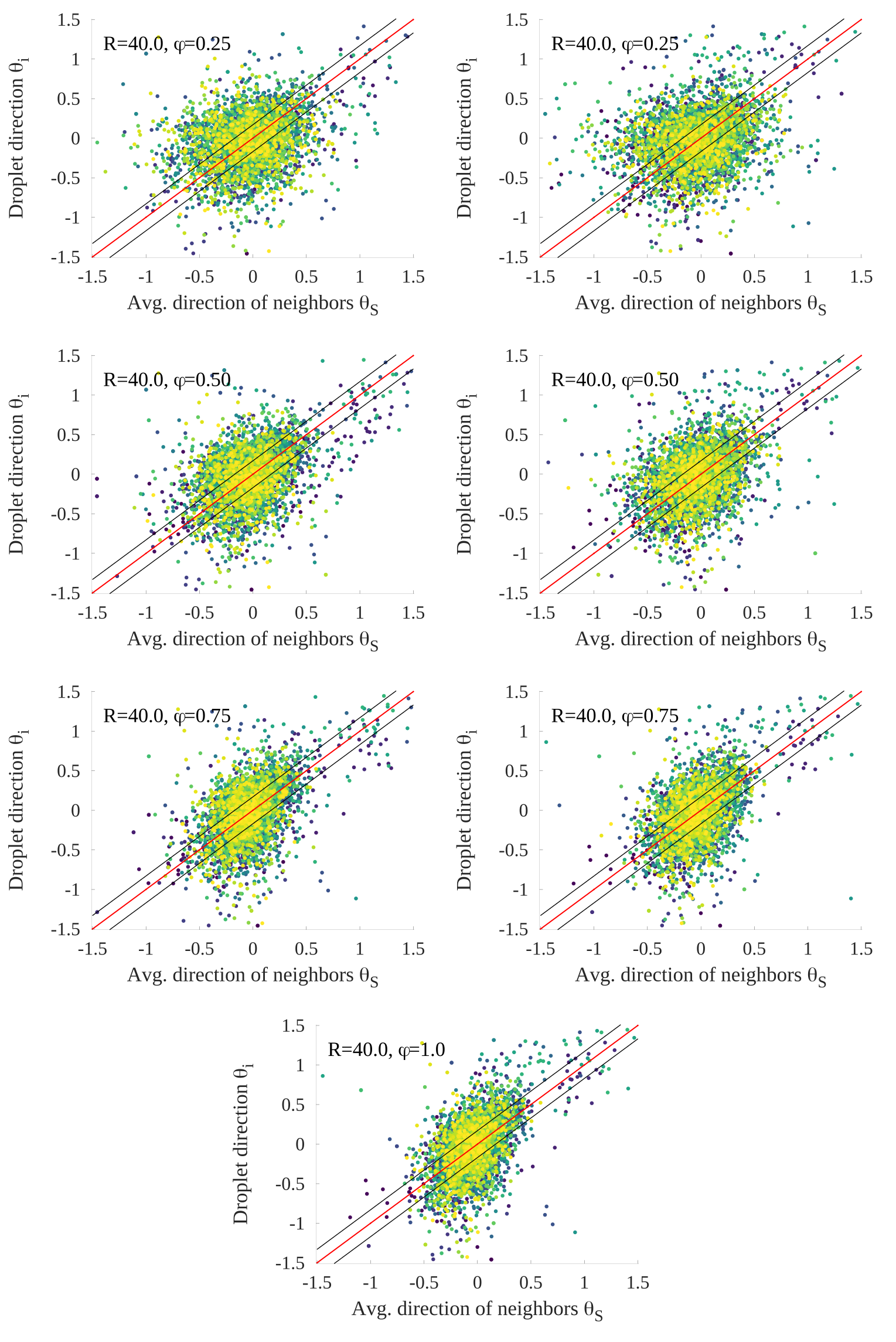}
\caption{Scatter plot of individual droplet's moving direction and average 
moving direction of its neighbors. The left column shows neighbor 
selection in the 
opposite direction of the motion of the droplet while the right column shows 
neighbors  selected in front of the moving droplet. The middle plot is computed 
with a full circle with radius $R$ as the neighborhood of a droplet. The range 
of $\phi$ is chosen as (0, 1), 
meaning the actual value of the angle is $2\pi\phi$. The points on the red line 
correspond to the instance when $\theta_i =\theta_S$, while points within the 
two black lines 
indicate instances where $\theta_i$ and $\theta_S$ are within 10 degrees of 
each other. Thus, points within the band bounded by two 
black lines show instances when individual droplet's 
moving direction is approximately same as it's neighbors. 
\label{scatter_plots}} 
\end{figure}

\begin{figure} []
\centering
\includegraphics[scale=0.45]{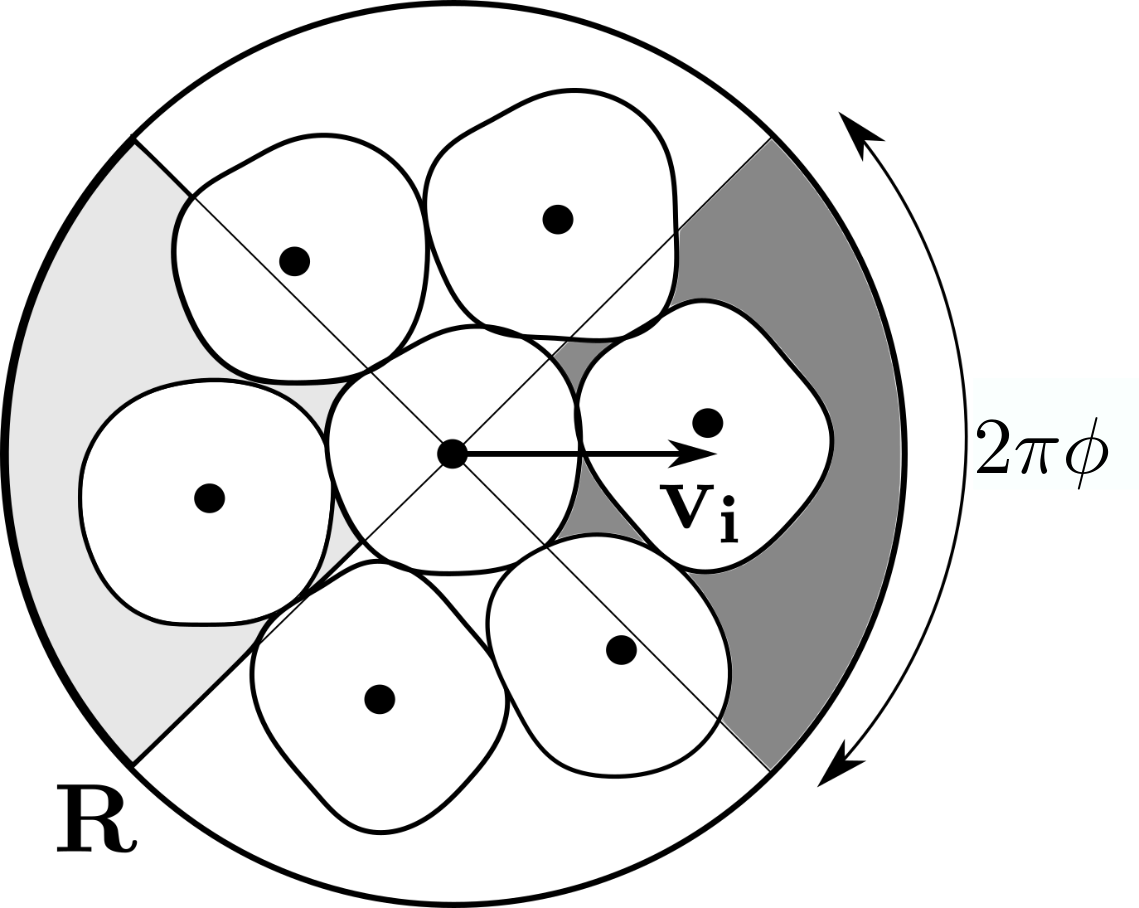}
\caption{The neighborhood of the droplet $i$, placed at the center of a circle 
of radius $R$, is the sector  of the circle enclosed by an angle $2\pi\phi$. 
The 
``front''  neighbors are the droplets whose center of mass is within 
the sector shaded with dark gray color while the 
``back'' neighbors are the ones whose center of mass 
is within the sector shaded with light gray color. The vector ${\bf v}_i$ is 
the speed of the droplet $i$.}\label{neighborhood} 
\end{figure}

\begin{figure} []
\centering
\includegraphics[scale=0.9]{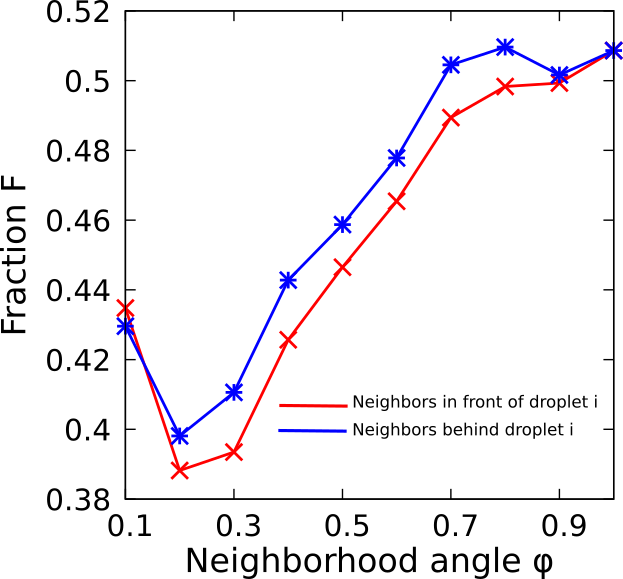}
\caption{Fraction of instances in which a given droplet is closely aligned with 
its neighbors as the neighborhood area is varied by changing the angle $\phi$. 
\label{analysis_scatter_plot}} 
\end{figure}

Figure~\ref{scatter_plots} shows a scatter-plot of 
an individual droplet's moving direction with the average direction of its 
neighbors. The neighbors of a droplet $i$ are defined as the droplets that are 
within distance $R$ from the droplet $i$ (see Fig.~\ref{neighborhood}) and 
within the sector of the circle with an angle $2\pi\phi$. The value of $R$ is 
chosen such that the circle contains the immediate neighbors of the droplet 
$i$, while the values of $\phi$ are varied between $0$ and $1$. We analyzed two 
cases, one in which the accounted neighbors are in front of the moving 
droplets, 
and a further one in which the neighbors are behind the moving droplets. This 
neighborhood definition allows us to see how these 
droplets positioned at two different locations affect the motion of an 
individual droplet $i$, an effect that can be inferred by counting the 
number of instances in which a given droplet is aligned with the average
direction of its neighbors. In Fig.~\ref{analysis_scatter_plot}, we plot
the fraction $F$ of points located  within a band indicated by the two black 
lines of each snapshot of Fig.~\ref{scatter_plots}. The points within this band 
correspond to an instance in which an individual droplet's moving direction is 
within ten degrees of its neighbors, i.e. $\vert \theta_i - \theta_S \vert< 
0.175$. Here $\theta_i$ is the moving direction of droplet $i$, and $\theta_S$ 
is the moving direction of neighbors of droplet $i$.
The results show that the droplets are aligned for a comparatively longer time 
with the droplets behind it rather than the droplets in front of it, thus 
suggesting that the droplets pushing from behind have higher influence on the 
individual droplet's moving direction.

\section{Conclusions}

In this work, we revisit the deep learning-based algorithms to infer the 
droplet trajectories by analyzing the output of lattice Boltzmann simulation
of dense emulsions. We measure various quantities like polar order parameter
and deviation of individual droplet's moving direction with its neighbors.  The 
results suggest that the droplets are aligned for a comparatively longer time 
with the droplets behind it rather than with  the droplets in front of it, 
meaning that the droplets pushing from 
behind have higher chance to affect the direction of an individual droplet.    
It would be interesting to perform similar analysis on data generated by active 
matter systems in order to further investigate to which extent the analogy with 
moving emulsions holds 
and provide further hints to write down a dynamical model of such emulsions in 
confined systems.

\section{Acknowledgments}
The authors acknowledge funding from the European Research Council under the 
European Union's Horizon 2020 Framework Programme (No. FP/2014-2020) ERC Grant 
Agreement No.739964 (COPMAT). We gratefully acknowledge the HPC 
infrastructure and the Support Team at Fondazione Istituto Italiano 
di Tecnologia.

\bibliography{Ref}
\bibliographystyle{unsrt}

\end{document}